\documentclass[aps,prl,
floatfix,twocolumn,showpacs,superscriptaddress]{revtex4-2}
\usepackage{graphicx}%
\usepackage{amsmath, amssymb}
\usepackage{color}
\usepackage{amsfonts}%Now
\usepackage{hhline}
\usepackage{braket}
\usepackage{bm}
\usepackage{mathrsfs}
\usepackage{hyperref}
\hypersetup{colorlinks=true,breaklinks,linkcolor=blue,urlcolor=blue,citecolor=blue}
\usepackage{xcolor}

\def\vec#1{\boldsymbol #1}
\usepackage{physics}

\def\be{\begin{equation}}
\def\ee{\end{equation}}
\def\bea{\begin{eqnarray}}
\def\eea{\end{eqnarray}}

\renewcommand{\vec}[1]{{\bm #1}}
\newcommand{\RNum}[1]{\uppercase\expandafter{\romannumeral #1\relax}}

\newcounter{pr}

\begin{document}

\title{
% Nonlinear Dynamics of Spiral Magnetic Order in the 1D-Biased Kondo Lattice Model
%Dynamics of Spin Spirals in a Biased Kondo Chain
% Dynamics of Spin Spirals under a Bias Voltage
%Bias-driven Dynamics of Spin Spirals in 1D Conductors 
Dynamics of spin spirals in a voltage biased 1D conductor
}

\author{Xiaohu Han}
\email{xhhan@csrc.ac.cn}
\affiliation{Beijing Computational Science Research Center, Beijing 100193, China}

\author{Pedro Ribeiro}
\email{pedrojgribeiro@tecnico.ulisboa.pt}
\affiliation{CeFEMA-LaPMET and Physics Department, Instituto Superior T\'{e}cnico, Universidade de Lisboa Av. Rovisco
Pais, 1049-001 Lisboa, Portugal}
\affiliation{Beijing Computational Science Research Center, Beijing 100193, China}

\author{Stefano Chesi}
\email{stefano.chesi@csrc.ac.cn}
\affiliation{Beijing Computational Science Research Center, Beijing 100193, China}
\affiliation{Department of Physics, Beijing Normal University, Beijing 100875, China}

\begin{abstract}
We analyze the fate of spiral order in a one-dimensional system of localized magnetic moments coupled to itinerant electrons under a voltage bias. Within an adiabatic approximation for the dynamics of the localized spins, and in the presence of a phenomenological damping term, we demonstrate the occurrence of various dynamical regimes: At small bias a rigidly rotating non-coplanar magnetic structure is realized which, by increasing the applied voltage, transitions to a quasi-periodic and, finally, fully chaotic evolution. These phases can be identified by transport measurements. In particular, the rigidly rotating state results in an average transfer of spin polarization. We analyze in detail the dependence of the rotation axis and frequency on system's parameters and show that the spin dynamics slows down in the thermodynamic limit, when a static conical state persists to arbitrarily long times. Our results suggest the possibility of discovering non-trivial dynamics in other symmetry-broken quantum states under bias.
%
%\pr{
%One-dimensional structures of localized magnetic moments coupled to metallic electrons can support ground states with magnetic spiral order.
%
%We investigate the behavior of this magnetic state under an applied bias voltage across the metal. We employ an adiabatic approximation with a phenomenological magnetic damping rate, assuming electronic time-scales are much faster than those of local moments. 
%
%We find that the magnetic order exhibits three different dynamic regimes, depending on the applied voltage and damping amplitudes. As voltage increases, the system transitions from a rigidly rotating magnetic structure to quasi-periodic dynamics, and finally to fully chaotic. These regimes can be identified by measuring the spin-polarized current. 
%
%We characterize the rigidly rotating regime, where the spin texture forms a deformed spiral with a damping-dependent rotation angle and frequency. At high damping rates, the rotation period increases faster than the system size which would be expected for extensive dissipation. Our results suggest the possibility of discovering non-trivial dynamics in other symmetry-broken quantum states under bias. }

\end{abstract}

\maketitle

{\it Introduction.---} The electric manipulation of spin polarization is a subject of great fundamental and practical interest. Notably, current-induced dynamics of domain walls and skyrmions in magnetic materials has been extensively explored, due to its promising applications in technologies like racetrack memories, magnetic transistors, or magnetic logic~\cite{RevModPhys.91.035004,PhysRevLett.93.127204,allwood2005magnetic, parkin2008magnetic, hayashi2008current, zhang2015magnetic}. These examples illustrate the strong influence that an applied bias can have on the properties of magnetic systems including, potentially, their phase diagram and critical behavior. Indeed, the general study of phase transitions of quantum systems under non-equilibrium conditions has recently gained increasing prominence (see, e.g., Refs.~\cite{2016RPP_Sieberer,2021RMP_dellatorre,2024arXiv_fazio}). For boundary-driven chains~\cite{2009JSMTE_PZ,2009PRB_benenti,2011PRZ_Znidarich,2013PRB_Clark,2014PRL_Prosen,PhysRevB.93.144305, PhysRevLett.122.235701,PhysRevB.103.035108,PhysRevB.107.125122,2022RMP_poletti}, it has been demonstrated that an external bias can induce exotic phase transitions between stationary states~\cite{PhysRevB.93.144305, PhysRevLett.122.235701,PhysRevB.103.035108,PhysRevB.107.125122}. Here, instead, we find that the equilibrium magnetic state can be destabilized by the applied bias, leading to a non-trivial collective dynamics. This behavior is reminiscent of a spontaneous breaking
of time-translation symmetry -- time-crystallinity -- allowed by the non-equilibrium conditions~\cite{PhysRevLettRoschAchim,PhysRevXRoschAchim}.

\begin{figure}%[!htbp]
			\centering
			\includegraphics[clip,width=0.47\textwidth]{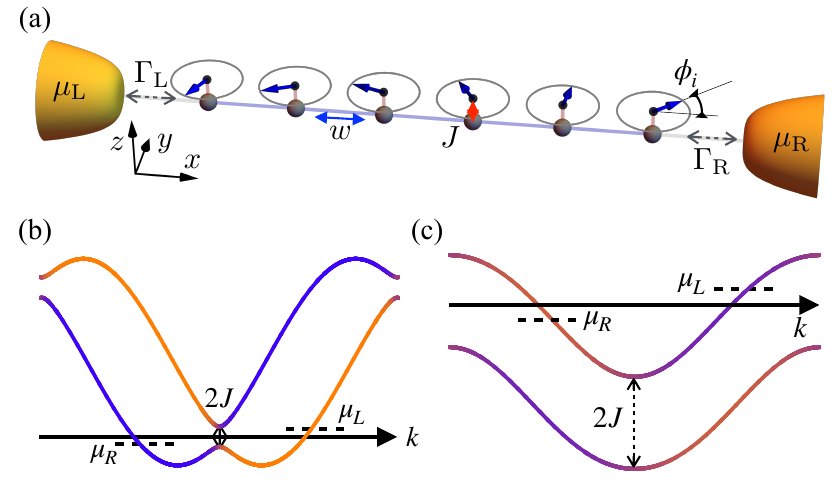}	
			\caption{(a): Lattice model of electrons coupled to two reservoirs. The blue arrows at the hopping sites are exchange-coupled localized spins $\bm m_i$.	(b) and (c): Electronic bands, obtained after a gauge transformation rotating $\bm m_i$ to the $x$ direction~\cite{SM, 2010PRB_Braunecker_Peierls, martin2012majorana, flensberg2015prl, hu2015interplay}. The color of the bands indicates $\langle \sigma_z \rangle$. In (b) and (c) we used $J=0.2w$ and $1.4w$, respectively.}
	\label{fig:fig1} %% label for entire figure
\end{figure}

Specifically, we consider the setup of Fig.~\ref{fig:fig1}(a), where classical spins are locally coupled to itinerant electrons. In equilibrium, the phase diagram of this system is notably rich, featuring ferromagnetic, antiferromagnetic, and more complex non-collinear magnetic states~\cite{flensberg2015prl, hu2015interplay,minami2015low, neuhaus2022complex}. Among them, spiral states have attracted particular interest~\cite{PhysRevB.80.165119, 2010PRB_Braunecker_Peierls,PhysRevLett.102.116403,flensberg2015prl, hu2015interplay,minami2015low, neuhaus2022complex}. At weak exchange coupling $J$, they can be understood to arise from the Ruderman-Kittel-Kasuya-Yosida (RKKY) interaction~\cite{RevModPhys.34.681, PhysRev.96.99, 10.1143/PTP.16.45, PhysRev.106.893}. Following a strong enhancement of RKKY coupling by electron-electron interaction, a spiral order of nuclear spins in nanowires has been predicted at accessible temperatures~\cite{PhysRevB.80.165119, 2010PRB_Braunecker_Peierls}, and experimental evidence of this phase has been found~\cite{PhysRevLett.102.116403}. The spiral state can also be realized~\cite{steinbrecher2018non} and locally probed with chains of magnetic atoms deposited on metallic surfaces~\cite{choi2019colloquium, heimes2015interplay}. Especially interesting is the chiral nature of this spiral state, which is of great relevance for topological superconductivity~\cite{kitaev2001unpaired, alicea2012new, leijnse2012introduction, martin2012majorana, braunecker2013interplay, klinovaja2013topological, vazifeh2013self, nadj2013proposal, li2014topological, nadj2014observation, kim2014helical, reis2014self, xiao2015chiral, pawlak2016probing, gorczyca2019topological, schecter2016self, pientka2013topological, christensen2016spiral, pientka2014unconventional, oreg2010helical, heimes2015interplay}. In proximity to a standard $s$-wave superconductor, such a system can host Majorana bound states, without requiring fine-tuning~\cite{braunecker2013interplay, klinovaja2013topological, vazifeh2013self}. 

In this letter we study the effects of an applied voltage on the symmetry-broken magnetic state, finding the onset of remarkable dynamical phenomena.
To illustrate how the voltage bias can destabilize the spiral order, consider in Figs.~\ref{fig:fig1}(b) and (c) a finite difference $\delta V = \mu_{L} -\mu_{R}$ in the chemical potentials of the chiral electronic states. The unbalanced occupation generates an electronic polarization along $z$ (i.e., perpendicular to the spiral plane) which, through the exchange interaction, drives a rigid precession of the localized spins. Furthermore, in the presence of suitable relaxation mechanisms, the spin texture deforms towards the $z$ direction. Since the spin evolution acts back on the electronic system, the bias can be expected to induce a dynamical interplay of spin and electron dynamics. Indeed, we find that a finite chain can realize magnetic textures rotating uniformly in time and, in the large bias regime, follows more complex types of chaotic or quasi-periodic time dependence. 

These dynamical states can be identified through measurements of the spin and charge current. They can also lead to transport of spin polarization, similar to a recently discussed magnetic Archimedes screw~\cite{Archimedeanscrew2021,Archimedeanscrew2022}. In that case, a uniform precession of the spiral is induced by an external oscillating magnetic drive~\cite{Archimedeanscrew2021} and results in spin pumping~\cite{Archimedeanscrew2022}. Here, an analogous periodic evolution is simply caused by the bias voltage, with interesting implication for the electric manipulation of the magnetization and spintronics applications in general.

{\it Model.---} The model is described by the Hamiltonian $ H=H_c+\sum_{l=\text{L},\text{R}}(H_l+H_{c,l}) $, where $H_c$ refers to the magnetic system. It reads  ($\hbar =1$):
\begin{equation} \label{Hc}
	H_c=- w \sum_{i,\sigma}  (c^\dagger_{i\sigma}c_{i+1\sigma}+{\rm H.c.})
    +2J \sum_i {\bm m}_i\cdot {\bm s}_i,
\end{equation}
where $w$ is the nearest-neighbor hopping amplitude and $J$ is the on-site exchange interaction between classical Heisenberg spins ${\bm m}_i$ and the electronic spin polarization, ${\bm s}_i = \frac12 \sum_{\sigma,\sigma'} \boldsymbol{\sigma}_{\sigma\sigma'} c^\dagger_{i\sigma}c_{i\sigma'}$. Here, $\boldsymbol{\sigma}$ is the vector of Pauli matrices and $ c^\dagger_{i\sigma} $ is the electron creation operator at site $i=1,2,\ldots L$, with spin index $\sigma =\uparrow , \downarrow $. For concreteness, we assume that the two reservoirs are semi-infinite tight-binding chains. For the right reservoir:
\begin{equation}
	H_\text{R}= -\sum_{i,\sigma}\left[w_\text{R} (c^\dagger_{i\sigma}c_{i+1\sigma}+{\rm H.c.})+\mu_\text{R} c^\dagger_{i\sigma}c_{i\sigma}\right],
\end{equation} 
where  $i=L+1,\ldots \infty$, and the corresponding coupling is $H_{c,\text{R}} = -\sum_{\sigma}w_{c-\text{R}}(c^\dagger_{L\sigma}c_{L+1\sigma}+{\rm H.c.})$. We define $H_{\text{L}}$ and $H_{c,\text{L}} $ in a completely analogous manner, with $i\leq 0$, intra-lead hopping amplitude  $w_\text{L}$, and coupling amplitude $w_{c-\text{L}}$ (between sites $i=0,1$). Furthermore, from the  two chemical potentials, we define $\delta V = \mu_\text{L}- \mu_\text{R}$ and $\mu = (\mu_\text{L}+\mu_\text{R})/2$.

{\it Methodology.---} We rely on the non-equilibrium Green's function formalism, under the simplifying assumption of wide-band reservoirs, i.e., $ w_{\rm L,R} $ are much larger than all other energy scales~\cite{PhysRevB.107.125122, PhysRevB.103.035108, PhysRevLett.122.235701}. Then, the coupling between the chain and the reservoirs is determined by the hybridization energies $ \Gamma_l=2\pi w_{c-l}^2\rho_{l} $, with $\rho_l$ the density of states of reservoir $l$.  Steady-state observables are easily computed from the single-particle correlation function $ \chi_{i\sigma,i'\sigma'} = \langle c_{i\sigma}c^\dagger_{i'\sigma'} \rangle$, and we can obtain $\bm \chi$ through the non-Hermitian single-particle operator $ \bm K=\bm H_c-i\sum_{l=\text{L},\text{R}}\bm{\gamma}_l/2 $, with ${\bm \gamma}_{l}=\Gamma_l \sum_\sigma c^\dagger_{r_l \sigma}  c_{r_l \sigma}$ acting on site $r_l=1,L$ for lead $l=\text{L},\text{R}$, respectively. By writing $\bm K=\sum_{\alpha}\lambda_\alpha\ket{\alpha}\bra{\alpha'}$, where $\ket{\alpha}$ and $\bra{\alpha'}$ are the right and left eigenvectors, respectively,  we can express $\bm \chi$ in the following form~\cite{PhysRevB.107.125122, PhysRevB.103.035108, PhysRevLett.122.235701}:
\begin{equation}\label{chi_explicit}
	\bm \chi=\frac{1}{2}\bigg(1+\sum_{l,\alpha, \beta} I_l(\lambda_\alpha,\lambda_\beta^*) \bra{\alpha'}{\bm \gamma}_l  \ket{\beta'} \ket{\alpha}\bra{\beta} \bigg) ,
\end{equation}
where $I_l(z,z')=\int{\frac{d\nu}{2\pi}\frac{\rm{tanh} \left [ \frac{\beta_l}{2}(\nu-\mu_l)\right ] }{(\nu-z)(\nu-z')}}$ with $ \beta_l=1/k_B T_l $. In this work we assume $T_{\text{L},\text{R}}=0$.

The above Eq.~(\ref{chi_explicit}) is derived for a fixed configuration of the classical spins ${\bm m}_i$, assuming that the itinerant electrons have reached the steady state. This occurs after a transient time of order $\tau_e$, which we suppose much shorter than the typical timescale $\tau_m$ for the evolution of ${\bm m}_i$. This adiabatic approximation is expected to hold in a broad range of parameters. In particular, a slow evolution of ${\bm m}_i$ is induced by a sufficiently small bias $\delta V$ and, as we will see, we find the divergent scaling relation $\tau_m \sim L^3$ in the intermediate and large damping regimes. 

In general, the electronic spin polarization  $ \left \langle{\vec s}_i\right \rangle $ from Eq.~(\ref{chi_explicit}) is not collinear to $\bm m_i$, resulting in a nontrivial torque $\propto  \left \langle{\vec s}_i\right \rangle \times \bm m_i $. 
%By including spin relaxation via a Gilbert damping term, we evolve the $\bm m_i$ according to:
We evolve the $\bm m_i$ as follows:
\begin{equation}\label{m_evolution}
\frac{\partial \vec{m}_i }{\partial t} =J\left \langle   \vec{{s}}_i \right \rangle\times \vec{m}_i+\eta (\left \langle   \vec{{s}}_i \right \rangle\times \vec{m}_i)\times \vec{m}_i,
\end{equation}
where dissipation is treated phenomenologically through the damping rate $\eta$. This effect is expected to be system-dependent and could arise from the electronic system itself, if treated beyond the adiabatic approximation~\cite{PhysRevLett.107.036804,PhysRevB.99.134409}. The dynamics described by Eq.~(\ref{m_evolution}) is highly non-linear, as the configuration of the ${\bm m}_i$ determines the electronic correlation function $\bm\chi$, which feeds back to Eq.~(\ref{m_evolution}) through $\left \langle   \vec{{s}}_i \right \rangle $.  

{\it Equilibrium state.---} Setting $\delta V=0$, we find the equilibrium configuration by evolving Eq.~(\ref{m_evolution}) to sufficiently long times~\cite{rackauckas2017differentialequations, SM}. This method represents an unbiased approach to the equilibrium configuration, which can realize complex forms of magnetic order~\cite{minami2015low,neuhaus2022complex}. In the Supplemental Information~\cite{SM}  we show that the correct spin configuration is recovered at various points of the phase diagram, including non-spiral states. Actually, the spiral order itself is only approximate, since a modulation of the relative angle at twice the Fermi wavevector of the chiral states persists in the thermodynamic limit~\cite{SM}. This weak residual instability of the spiral states was not recognized in previous studies, but has limited consequences on our following discussions, except that it complicates the finite-size scaling analysis~\cite{SM}.

\begin{figure}%[!htbp]
			\centering
			\includegraphics[clip, width=0.47\textwidth]{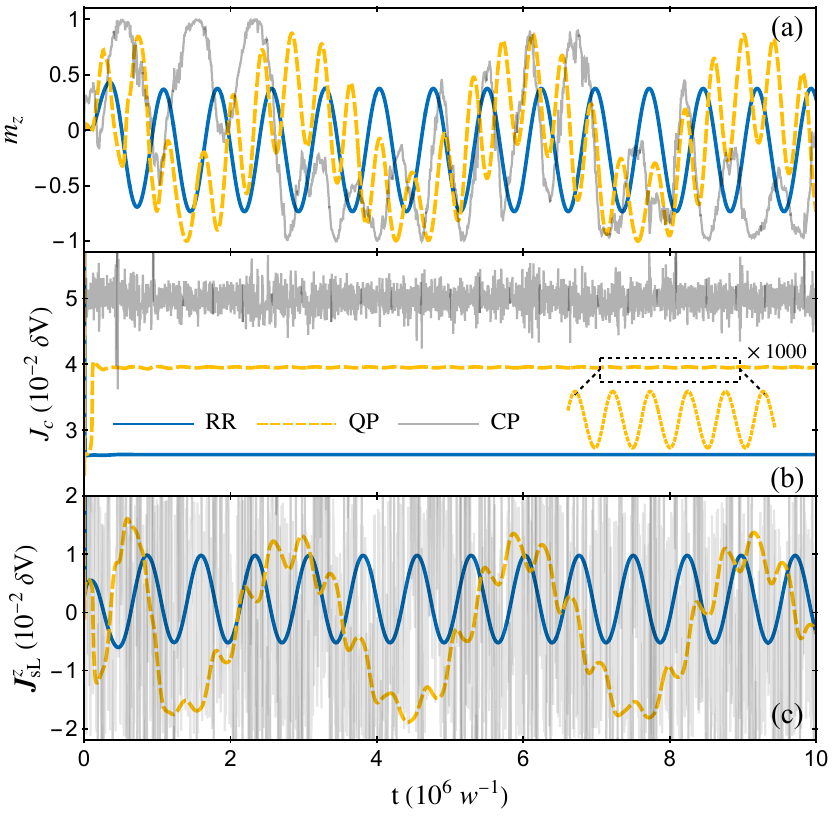}
	\caption{Comparison between RR, QP and CP dynamics. We show: (a) evolution of $m_z$ for one of the two middle sites, (b) charge current, and	(c) $z$-component of the spin current.  In all panels we used $L=500$, $J/w=1.4$, $\mu/w=1.5$, $ \eta/w =0.2 $, and $ \delta V/w=0.03 $ 
 (RR), $ 0.033$ (QP), and $0.08 $ (CP). }
	\label{fig:comparison} %% label for entire figure
\end{figure}

{\it Non-equilibrium dynamics.---} Applying a finite bias drives the system away from equilibrium, inducing a complex dynamics of the classical spins. The system has no steady state, since setting $\partial \bm m_i/\partial t = 0$ makes it impossible to solve Eq.~(\ref{m_evolution}). Instead, the ansatz:
\begin{equation}
\label{eqn:rrEOM}
\frac{\partial \vec{m}_i }{\partial t} =\vec \Omega \times \vec{m}_i,
\end{equation}
allows us to find suitable values of $\bm \Omega$ and $\bm m_i$, corresponding to a rigidly rotating (RR) magnetic state. An explicit numerical integration of Eq.~(\ref{m_evolution}) is shown in Fig.~\ref{fig:comparison} where, for definiteness, we take $J=1.4w$ and $\mu/w=1.5$. At this point of the phase diagram an equilibrium spiral state $\bm m_i=(\cos q_si,\sin q_si,0)$ is found~\cite{hu2015interplay, neuhaus2022complex}, with $q_s \simeq \pi/4$, and the corresponding bands are plotted in Fig.~\ref{fig:fig1}(c). As shown in Fig.~\ref{fig:comparison}(a), at small voltage the long-time dynamics follows the periodic evolution predicted by Eq.~(\ref{eqn:rrEOM}). Instead, at larger values of $\delta V$, the RR solution is not a stable limit cycle, and quasi-periodic (QP) dynamics appears. Eventually, a chaotic phase (CP) is realized at larger values of $\delta V$.  With smaller values of $J$, e.g., like in Fig.~\ref{fig:fig1}(b), we obtain results equivalent to Fig.~\ref{fig:comparison}~\cite{SM}. 

If the magnetization dynamics is not directly accessible, the three phases can be distinguished in transport. Consistently with our adiabatic approximation, we compute the instantaneous electron current $J_c = 2w\sum_{\sigma} {\rm Im}(\langle c_{i\sigma}c^\dagger_{i+1\sigma} \rangle )$ (independent of $i$), plotted in Fig.~\ref{fig:comparison}(b). Since a global rotation of the $\bm m_i$ does not affect the charge properties, the current approaches a constant in the RR regime. Instead, the presence of two incommensurate frequencies results in an oscillatory component for the QP solution. Finally, the chaotic evolution is obvious from $J_c$. 

The RR state can be distinguished from a stationary configuration by probing the spin current. In Fig.~\ref{fig:comparison}(c) we have computed
$\vec J_{s \text{L}}= - \partial_t \sum_{i\leq 0}\sum_{\sigma\sigma'} \langle c^\dagger_{i\sigma}\vec{\sigma}_{\sigma\sigma'} c_{i\sigma'}  \rangle$, which is the rate of spin polarization drained from the left reservoir. As expected, we find an oscillatory dependence for the RR state and, interestingly, $J_{s \text{L}}^z$ has a finite time average. In general, $ \bar{\vec J}_{s \text{L}}=\int_{T_1}^{T_1+NT}\vec J_{s \text{L}}/NT\propto \vec \Omega $, since the spin current vector is rotating in time with the magnetic configuration. Similar results hold for $\vec J_{s \text{R}}$, indicating that the RR states induce a transfer of spin polarization, oriented along the rotation axis.

\begin{figure}%[!htbp]
			\centering
			\includegraphics[width=0.47\textwidth]{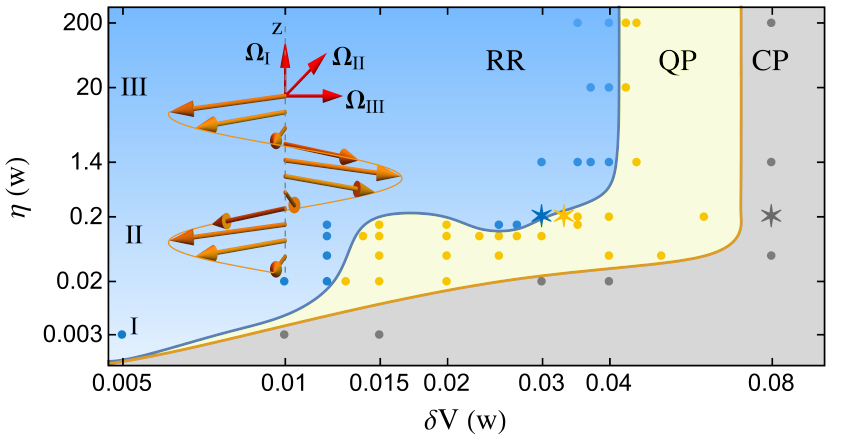}
	\caption{
	Non-equilibrium phase diagram for $J/w=1.4$, $\mu/w=1.5$, and $ L=500 $. Solid dots mark the states which were explicitly classified, and the three stars correspond to Fig.~\ref{fig:comparison}. The schematics indicates the orientation of $\bm \Omega$ with respect to the plane of the spiral when $\delta V\to 0$ and $\eta$ is small (I), intermediate (II), or large (III).  
	} 
	\label{fig:phase_diagram} %% label for entire figure
\end{figure}

{\it Phase diagram.---} We have shown that three dynamical regimes can be distinguished by their transport properties. The resulting phase diagram, in terms of $\delta V$ and $\eta$, is shown in Fig.~\ref{fig:phase_diagram}. Although the phase boundaries depend on $L$, the results of Fig.~\ref{fig:phase_diagram} (where  $L=500$) are representative of the thermodynamic limit~\cite{SM}. We find that the voltage leading to a breakdown of the RR phase has a sensitive dependence on $\eta$ only when $\eta \lesssim w$. In this regime, the size of the QP region is also strongly affected by $\eta$. We also find an interesting re-entrant behavior of RR as function of $\delta V$ around $\eta/w \sim 0.15$. Due to the diverging relaxation time, the asymptotic evolution when $\eta \to 0$ is more difficult to characterize, and we have not reached definite conclusions about the existence of an intermediate QP region in this limit.

\begin{figure}%[!htbp]
		\centering
		\includegraphics[clip, width=0.47\textwidth]{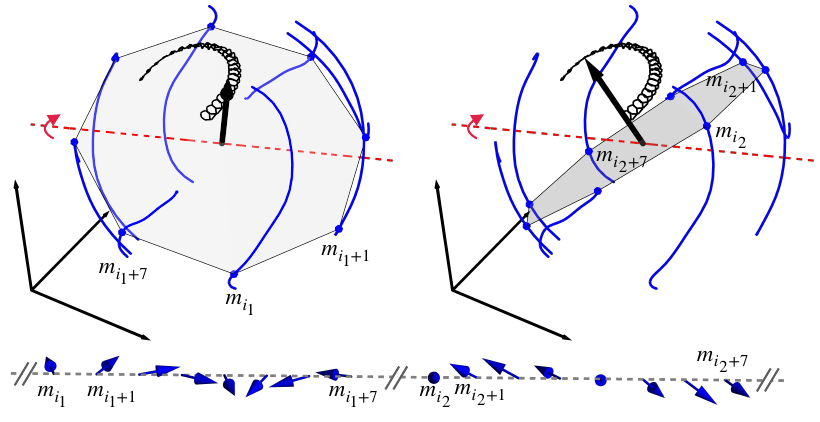}
	\caption{
	Classical spin configuration in a RR state with large bias. In each of the upper plots, the blue lines connect the orientations of 200 middle spins $\bm m_{i}$, brought to a common origin ($L=500$ and $150\leq i < 350$). In the left plot, we have marked with blue dots the spins with $i=200,~201,\ldots 207$ ($i_1=200$). In the right plot, we mark eight consecutive spins with $i_2=320$. In both cases, the black arrows are normal vectors to the gray planes of eight neighboring spins, the black lines trace the normal vector along the chain, and the red dashed line indicates the rotating axis. In the bottom part of the figure we represent the spin chain, considering the same two sets of eight spins. Other parameters correspond to the RR state marked with a blue star in Fig.~\ref{fig:phase_diagram}.
	 }
	\label{fig:Deforming} %% label for entire figure
\end{figure}

While in the QP and CP phases the relative orientations of the $\bm m_i$ change in time, the RR states are still characterized by well-defined magnetic configurations, with fixed relative orientations of the $\bm m_i$. In the limit of small bias, the $\bm m_i$ remain close to the planar equilibrium state and, as illustrated in the inset of Fig.~\ref{fig:phase_diagram}, we can distinguish three regimes. For $\eta \ll J $ (region I), the precession vector $\bm \Omega$ is perpendicular to the plane of the spiral. Instead, $\bm \Omega$ lies in the plane when $\eta \gtrsim  J$ (region III). Finally, $\bm \Omega$ has a nontrivial orientation in the crossover region II. At large values of $\delta V$ these three regimes are not well-defined, since the classical spins strongly deviate from the equilibrium configuration. An example is shown in Fig.~\ref{fig:Deforming}: We see that locally the $\bm m_i$ are close to a spiral state. However, the approximate spiral plane changes along the chain, resulting in a nontrivial 3D configuration. 

{\it Linear response regime.---} In the near-equilibrium limit, the system is always in the RR phase. We can then transform Eq.~(\ref{m_evolution}) to the appropriate rotating frame, leading to a stationary spin configuration $\widetilde{\bm{m}}_i$ and electronic polarizations $\langle \widetilde{\vec{s}}_i\rangle$. The angular velocity is $\bm \Omega \simeq  \bm \omega \delta V$ and the spin configuration is close to an equilibrium solution, thus $\widetilde{\vec{m}}_{i}\simeq \vec{m}_{i}^{eq}+\vec{n}_{i} \delta V $. Similarly, we expand:
\begin{equation}
\langle \widetilde{\vec{s}}_i\rangle \simeq \langle \vec{s}_i\rangle^{eq}+\left(\left.\frac{\partial \langle \widetilde{\vec{s}}_i\rangle}{\partial  \delta V}\right|_{eq}+\sum_{j=1}^L \left. \frac{\partial \langle \widetilde{\vec{s}}_i\rangle}{\partial \widetilde{\vec{m}}_{j}} \right|_{eq}\cdot \vec{n}_{j} \right) \delta V.
\end{equation}
Since $\langle \widetilde{\vec{s}}_i \rangle$ can be numerically computed at given $\delta V$ and $\widetilde{\bm m}_i$ using Eq.~(\ref{chi_explicit}), its partial derivatives at $\delta V=0$ and $\widetilde{\bm m}_i = \vec{m}_{i}^{eq}$ can be obtained explicitly. The only variables of the linearized equations are $\bm \omega$ and $\vec{n}_{i}$, satisfying:
\begin{align}\label{linearized_eq}
\left[
 J \bm A + \eta ( \bm A \times   \bm m^{eq}_i )
 -{\bm \omega}
\right]  
\times
\bm{m}^{eq}_i
= 0
,
\end{align}
where $\bm A = |\langle \vec{s}_i\rangle^{eq}| \vec{n}_{i} +
\sum_{j=1}^L   \left. \frac{\partial \langle \widetilde{\vec{s}}_i\rangle}{\partial \widetilde{\vec{m}}_{j}} \right|_{eq}\cdot \vec{n}_{j}
 +\left.\frac{\partial \langle \widetilde{\vec{s}}_i\rangle}{\partial  \delta V}\right|_{eq}$.
%\begin{align}\label{linearized_eq}
%	&  
%\bm B_i\left(
%|\langle \vec{s}_i\rangle^{eq}| \widetilde{\vec{m}}_{i} +
%\sum_{j=1}^L   \left. \frac{\partial \langle \vec{s}_i\rangle}{\partial \vec{m}_{j}} \right|_{eq}\cdot \widetilde{\vec{m}}_{j}
 %+\left.\frac{\partial \langle \vec{s}_i\rangle}{\partial  \delta V}\right|_{eq}
%\right)  
%\times
%\bm{m}^{eq}_i\nonumber\\
%& 
%= \widetilde{\bm \Omega}\times \bm{m}^{eq}_i
%,
%\end{align}
%where we defined $\bm {B}_i(\bm A) = J \bm A + \eta ( \bm A \times   \bm m^{eq}_i )$. 
To derive Eq.~(\ref{linearized_eq}), we neglected higher-order corrections and used $\langle \vec{s}_i\rangle^{eq}=-|\langle \vec{s}_i\rangle^{eq}|\bm{m}^{eq}_i $, i.e., in equilibrium the classical spins are anti-parallel  to $\langle\vec{s}_i\rangle^{eq}$.

\begin{figure}%[!htbp]
			\centering
			\includegraphics[clip, width=0.47\textwidth]{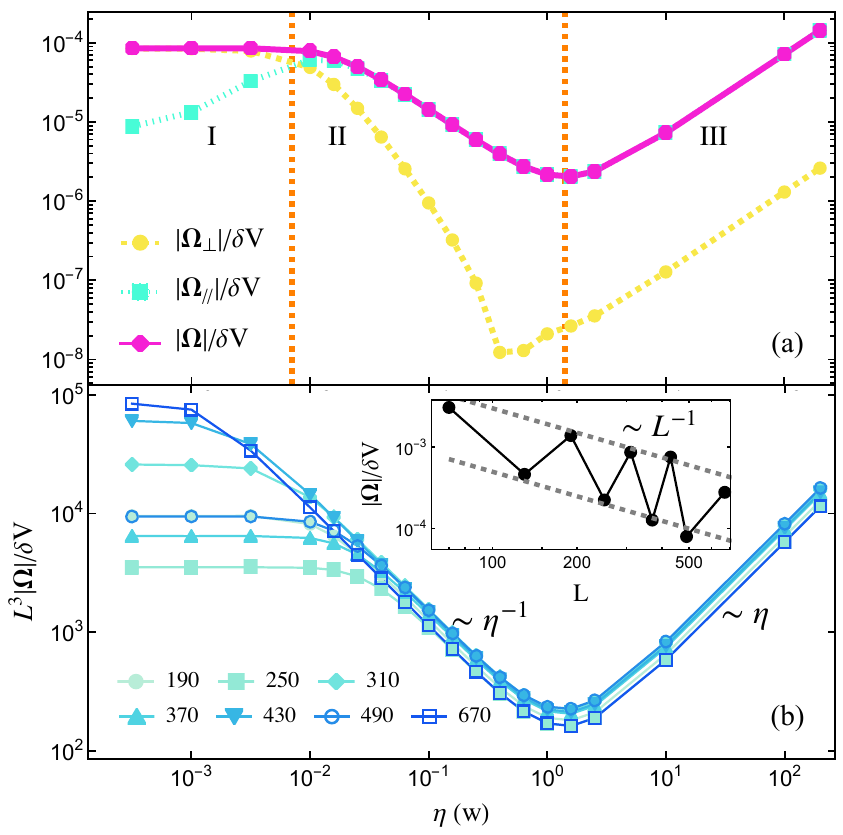}
	\caption{Dependence on $\eta$ of the precession vector $ \vec\Omega $, in the limit $ \delta V \to 0$. Panel (a) shows how the orientation of $ \vec\Omega $ (with $L=500$) changes from perpendicular (I), to tilted (II) and, finally, parallel (III) to the spiral plane. Panel (b) shows that $ |\vec\Omega| \propto L^{-3}$ in region II and III. Instead, $ |\vec\Omega| \propto L^{-1} $ in region I, with large finite size effects (see inset).
	 }
	\label{fig:scaling} %% label for entire figure
\end{figure}

The dependence of $\bm \Omega$ on $\eta$ at a given system size, obtained from Eq.~(\ref{linearized_eq}), is shown in Fig.~\ref{fig:scaling}(a). To highlight the three regimes illustrated in Fig.~\ref{fig:phase_diagram}, we separate $ \vec\Omega=\vec\Omega_{\parallel}+\vec\Omega_\bot $, where $\vec\Omega_{\parallel}$ ($\vec\Omega_\bot$) is coplanar (perpendicular) to the $\bm{m}^{eq}_i$. We find $|\vec\Omega_{\parallel}|\ll |\vec\Omega_\bot|$ in region I, when the precessing spins remain in the same plane of the equilibrium configuration, and $|\vec\Omega_{\parallel}|\gg |\vec\Omega_\bot|$ in region III. The three regimes are also clearly distinguished by the dependence of the precession frequency on $\eta$: $|\vec\Omega|$ is approximately constant in region I, while $|\vec\Omega|\propto \eta^{-1}$ in region II and $|\vec\Omega|\propto \eta$ in region III.

Finally, we study in Fig.~\ref{fig:scaling}(b) the finite-size scaling of the precession frequency $|\vec\Omega|$. In regions II and III we can collapse the data by assuming a $|\vec\Omega| \propto L^3$ dependence. The scaling properties at $\eta \to 0$ are difficult to ascertain, due to the long relaxation time and large finite-size effects, but our numerical data suggest that $|\vec\Omega| \propto L$  (see inset). In summary, we find the following approximate dependencies on $\eta$ and $L$: 
\begin{equation}
	\frac{|\vec\Omega|}{\delta V} \propto 
\left\{
\begin{array}{ll}
L^{-1}  & {\rm in~region~I},\\
\eta^{-1} L^{-3}  & {\rm in~region~II},\\
\eta L^{-3}  & {\rm in~region~III}.\\
\end{array}
\right.
\end{equation}
In Fig.~\ref{fig:scaling}(a) we see that the minimum of $|\bm \Omega|$, which separates regions~II and~III, occurs quite accurately at $\eta = J$. Instead, due to the different scaling laws, the crossover between regions~I and II occurs at $ \eta \propto L^{-2}$. The shrinking of region I with $L$ can also be seen in Fig.~\ref{fig:scaling}(b), and we expect that for $L\to \infty$ the dependence $L^3|\bm \Omega|/\delta V \propto \eta^{-1}$ remains valid down to vanishing values of $\eta$.

\begin{figure}%[!htbp]
			\centering
			\includegraphics[width=0.47\textwidth]{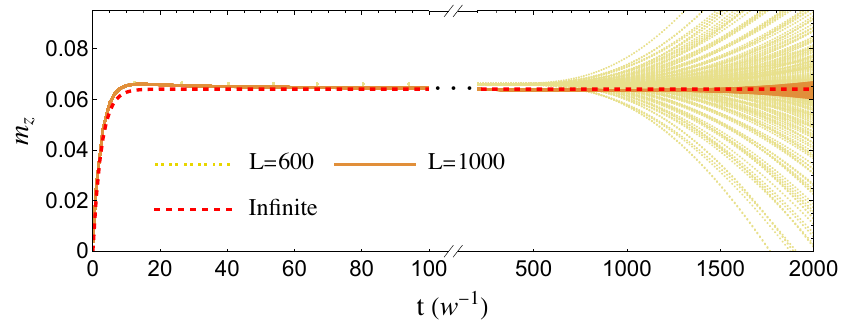}
	\caption{
	Time dependence of $m_{i,z}$ for the middle $ 40\% $ classical spins of a chain with $L=600$ (yellow) and $L=1000$ (orange). The red dashed curve is obtained from a tilted-spiral ansatz. Significant deviations from this approximate dynamics start to appear at $t_1 \simeq 500$  for $L=600$ and $t_1\simeq 1500$ for $L=1000$. Here we used $\eta/w=200 $ and $ \delta V/w=0.03 $.
	}
\label{fig:InfiniteSystem} %% label for entire figure
\end{figure}

{\it Infinite system.---} The above scaling laws imply that the dynamics of RR gets frozen in the thermodynamic limit (since $|\bm \Omega| \to 0$). Therefore, if we consider $L\to \infty$ \emph{before} taking $t \to \infty$, it becomes necessary to examine the transient evolution, taking place before the RR state is established. This short-time dynamics is shown in Fig.~\ref{fig:InfiniteSystem} where, interestingly, all the bulk classical spins follow a nearly equivalent evolution. They are well described by the ansatz  $\bm m_i=(m_\parallel \cos (q_s i+\varphi),m_\parallel \sin (q_s i+\varphi), m_z)$, with time-dependent values of $m_z$ and $\varphi$. The system is quickly driven to a stationary conical state, where the classical spins acquire a uniform out-of-plane component. As seen in  Fig.~\ref{fig:InfiniteSystem}, this type of evolution is almost independent of system size, and lasts until a timescale which diverges with $L$~\cite{SM}. 

This behavior can be well understood from the infinite-system limit by applying a gauge transformation, which brings the conical states (including the spiral~\cite{flensberg2015prl, hu2015interplay}) to uniform configurations. In the transformed frame, the spin polarization $\langle \bm s_i \rangle$ is uniform as well, thus all the classical spins follow an identical time evolution. $\langle \bm s_i \rangle$ is determined by non-equilibrium distribution functions $n_{\alpha}(k)$ of the $(\alpha,k)$ Bloch states~\cite{SM} which, as described in more detail in the Supplemental Information~\cite{SM}, we have extracted numerically. The result of this approximate treatment, shown by the red dashed curve of Fig.~\ref{fig:InfiniteSystem}, is in good agreement with the finite-system simulations.  

{\it Conclusion.---} In summary, we have shown that even an infinitesimal bias can induce non-trivial dynamics in states with symmetry-broken spiral order. For small biases, the spiral structure rotates rigidly with a precession frequency that decreases with system size. While in the weak damping regime the precession vector is perpendicular to the spiral plane, it crosses-over to be parallel to the plane for strong damping. Upon increasing the bias voltage, the rigidly rotating spiral structure gets deformed into a non-trivial 3D configuration, before transitioning to a non-rigid rotating quasi-periodic phase and further switching to a chaotic regime at even higher voltages. These different dynamical states can be identified through measurements of the spin and charge current.

Our work suggests that non-equilibrium conditions imposed by reservoirs can naturally induce dynamics in the Goldstone mode of symmetry-broken phases. These results suggest the possibility of discovering non-trivial dynamical phases in other symmetry-broken quantum states under bias. Understanding under which conditions they may arise is an interesting direction for future research. Direct extensions of our analysis may consider the consequences of superconductivity, inducing a topological regime, or electron interactions. 

\begin{acknowledgments}
We thank S. Kirchner, P. Liang, and A. Russomanno for useful discussions. This research is supported by FCT-Portugal (PR), as part of the QuantERA II project “DQUANT: A Dissipative Quantum Chaos perspective on Near-Term Quantum Computing” via Grant Agreement No. 101017733. PR acknowledges further support from FCT-Portugal through Grant No. UID/CTM/04540/2020. S.C. acknowledges support from the Innovation Program for Quantum Science and Technology (Grant No. 2021ZD0301602) and the National Science Association Funds (Grant No. U2230402).

\end{acknowledgments}

\bibliography{nonequ.bib}

\newpage 
\clearpage\thispagestyle{empty}

\setcounter{section}{0}
\setcounter{figure}{0}

\renewcommand{\theequation}{S\arabic{equation}}
\renewcommand{\thefigure}{S\arabic{figure}}
\renewcommand{\thetable}{S\arabic{table}}
\renewcommand{\thesection}{S\arabic{section}}

\onecolumngrid

\begin{center}
       %\vspace*{0.4cm}opo
        \Large{Supplemental Material for ``Dynamics of spin spirals in a voltage biased 1D conductor''}
        
        \vspace{0.4cm}
        \normalsize
        Xiaohu Han,$^1$ Pedro Ribeiro,$^{2,1}$ and Stefano Chesi$^{1,3}$
        \\
        \textit{
        $^1$Beijing Computational Science Research Center, Beijing 100193, China\\
        $^2$CeFEMA-LaPMET and Physics Department, Instituto Superior T\'{e}cnico, \\ Universidade de Lisboa Av. Rovisco
Pais, 1049-001 Lisboa, Portugal \\
        $^3$Department of Physics, Beijing Normal University, Beijing 100875, China} 
		%\\ \vspace{0.2cm}
		%(Dated: \today) 
       \vspace{0.5cm}
\end{center}

\twocolumngrid

\section{Electronic bands \label{Sec:bands}}

In this Section we compute the electronic states by assuming in Eq.~(\ref{Hc}) a tilted spiral configuration:
\begin{equation}\label{tilted_spiral}
\bm m_i=(m_\parallel \cos (q_si+\varphi), m_\parallel \sin (q_si+\varphi),m_z),
\end{equation}
where $m_\parallel = \sqrt{1-m_z^2}$.  By applying the gauge transformation~\cite{2010PRB_Braunecker_Peierls, martin2012majorana, flensberg2015prl, hu2015interplay}
\begin{align}\label{GTransf}
c^\dagger_{j\uparrow} \rightarrow c^\dagger_{j\uparrow}e^{iq_sj/2}, \quad
c^\dagger_{j\downarrow} \rightarrow c^\dagger_{j\downarrow}e^{-iq_sj/2},
\end{align}
the Hamiltonian becomes:
\begin{align}\label{tildeHc}
\bar{H}_c= & - \sum_{i} w(e^{-iq_s/2}c^\dagger_{i\uparrow}c_{i+1\uparrow}
+e^{iq_s/2}c^\dagger_{i\downarrow}c_{i+1\downarrow} + {\rm H.c.}) \nonumber \\ 
&+ 2J \sum_i \bar{\bm m}_i\cdot {\bm s}_i,
\end{align}
where the classical spins now have a uniform orientation: 
\begin{equation}
\bar{\bm m}_i=(m_\parallel\cos\varphi,m_\parallel\sin\varphi,m_z).
\end{equation}
Applying $\hat{c}_{j\alpha}^{\dagger}=\frac{1}{\sqrt{L}} \sum_{k} e^{-i k  j} \hat{c}_{k\alpha}^{\dagger}$ gives:
\begin{equation}
%\tilde{H}_c=\sum_{k}\left ( c_{k\uparrow} ^\dagger ,c_{k\downarrow } ^\dagger  \right ) \left ( \begin{matrix}
% \varepsilon_{k-q_s/2} &J \\ 
%J  & \varepsilon_{k+q_s/2}
%\end{matrix} \right ) \binom{c_{k\uparrow}}{c_{k\downarrow }} ,
\bar{H}_c=\sum_{k}\left ( c_{k\uparrow} ^\dagger ,c_{k\downarrow } ^\dagger  \right )
\bm{H_0}
\binom{c_{k\uparrow}}{c_{k\downarrow }} ,
\end{equation}
where
\begin{equation}\label{H0_matrix}
\bm{H_0}=\left ( \begin{matrix}
 \varepsilon_{k-q_s/2} + Jm_z &Jm_\parallel \rm{e}^{-i\varphi} \\ 
Jm_\parallel \rm{e}^{i\varphi}  & \varepsilon_{k+q_s/2} - Jm_z
\end{matrix} \right ),
%\bm{H_0}=\left ( \begin{matrix}
% -2 w \cos{k-q_s/2} + Jm_z &Jm_\parallel \\ 
%Jm_\parallel  & -2 w \cos{k+q_s/2} - Jm_z
%\end{matrix} \right )
\end{equation}
and $\varepsilon_k = -2 w \cos k $. 
$\bar{H}_c$ can now be readily diagonalized:
\begin{equation}\label{Epmk}
\bar H_c = \sum_{\alpha =\pm} E_\alpha(k) d_{k \alpha}^\dag  d_{k \alpha}.
\end{equation} 
In the planar planar case, i.e. $m_z=0$:
\begin{align}
 E_{\pm}	(k)= & -w\cos(k+\frac{q_s}{2})-w\cos(k-\frac{q_s}{2}) \nonumber \\
&\pm \sqrt{J^2+ w^2 \left[\cos\left(k+\frac{q_s}{2}\right)-\cos\left(k-\frac{q_s}{2}\right) \right]^2}.
\end{align}
The above expression is used to plot Figs.~\ref{fig:fig1}(b) and (c).

\section{Equilibrium state}
%\label{sec:EquilibriumMagneticConfigurations}
Solving Eq.~(\ref{m_evolution}) with $\delta V=0$ is the starting point of our analysis and serves as an important benchmark of our method. To obtain the equilibrium state, we take a random configuration of the classical spins $\bm m_i$ and evolve the system for sufficiently long times. A stationary configuration is reached when each $\bm m_i$ is antiparallel to its on-site electron polarization $\left \langle{\vec s}_i\right \rangle$. In practice, we stop the evolution when the condition ${\rm max}_i 
\left|\cos^{-1}
\frac{{\bm m}_i \cdot \left \langle{\vec s}_i\right \rangle}
{\left|\left \langle{\vec s}_i\right \rangle\right|}-\pi
\right|<10^{-6}$ is satisfied. The final configurations are generally robust with respect to the choice of the random initial state.

\begin{figure}%[!htbp]
			\centering
			\includegraphics[width=0.47\textwidth]{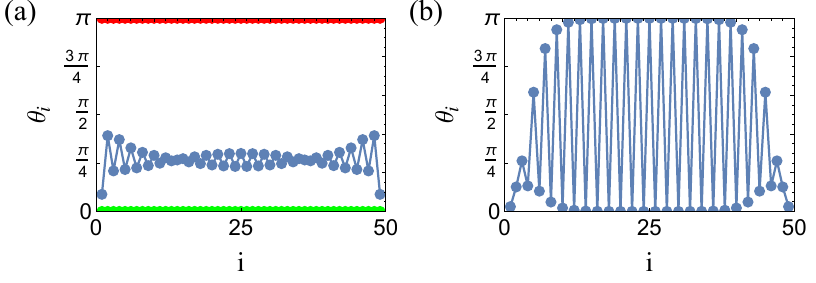}
	\caption{
	Equilibrium configurations at different values of $J$ and $\mu$, described by the relative angle $\theta_i$ between neighboring spins, see Eq.~(\ref{relative_angle}). In panel (a) we show states well described by the spiral ansatz. The green curve (FM order) is at $ J/w=1.0$ and $\mu/w=1.9$. The red curve (AF order) is at $ J/w=1.4$ and $\mu/w=0.5 $. The blue curve (spiral, with $q_s \simeq \pi/4$) is at  $ J/w=1.4$ and $\mu/w=1.5 $. The state of panel (b) is at $ J/w=1.0$ and $\mu/w=1.5 $, giving $\uparrow \uparrow \downarrow \downarrow$ order.
	}
\label{fig:figAppen1} %% label for entire figure
\end{figure}

In Fig.~\ref{fig:figAppen1} we show representative results obtained with this approach. The phase diagram has been studied in detail by Refs.~\cite{minami2015low, hu2015interplay, neuhaus2022complex}, and admits both ferromagnetic (FM) and anti-ferromagnetic~(AF) regions as the two limits of the planar spiral state  ($q_s=0$ and $\pi$, respectively, while $m_z=0$). Furthermore, complex types of spin order have been found in certain portions of the phase diagram~\cite{minami2015low, neuhaus2022complex}. In agreement with Ref.~\cite{neuhaus2022complex}, we only find planar states, thus we may choose ${\bm m}_i = (\cos\phi_i ,\sin \phi_i,0)$. The spin configuration is specified by the relative angles:
\begin{equation}\label{relative_angle}
\theta_i = \phi_{i+1}-\phi_i,
\end{equation}
which are plotted in Fig.~\ref{fig:figAppen1} for different choices of $J$ and $\mu_{\rm L}=\mu_{\rm R}\equiv \mu$. The spiral ansatz is $\phi_i = q_s i$, giving a constant value $\theta_i = q_s$. This type of state is approximately realized by the $q_s \simeq \pi/4$ configuration shown in panel~(a). In Fig.~\ref{fig:figAppen1}(a) we also show instances of FM and AF order, found at values of $J$ and $\mu$ in agreement with previous studies. Furthermore, we have considered a point in the phase diagram where direct minimization of the free energy leads to a collinear state of the form $\uparrow \uparrow \downarrow \downarrow$~\cite{neuhaus2022complex}. As seen in Fig.~\ref{fig:figAppen1}(b), evolving Eq.~(\ref{m_evolution}) leads to the correct spin configuration in the bulk of the system.

\begin{figure}%[!htbp]
			\centering
			\includegraphics[clip, width=0.47\textwidth]{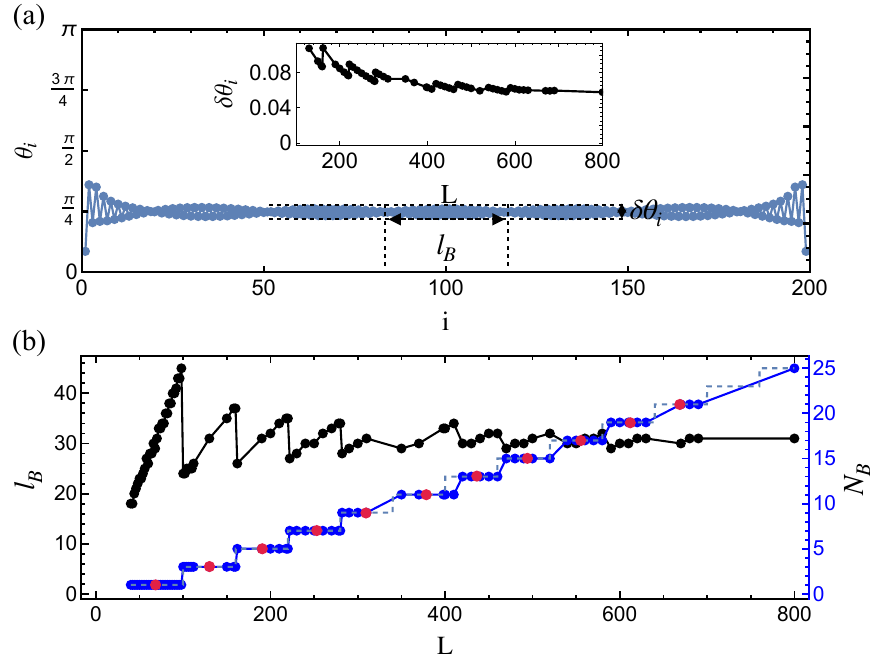}
	\caption{
	(a) Modulation of the equilibrium spiral configuration. For the middle 100 spins of the chain, we define $\delta \theta_i = (\theta_i)_{max}-(\theta_i)_{min}$. The inset shows the dependence of $\delta \theta_i$ on $L$.
	(b) Length $l_B$ of the beating period and number $N_B$ of beating periods, as function of $L$. Red dots mark the center of each interval. We used $J/w=1.4$ and $\mu/w = 1.5$.
	%\color{red} Inset: $q_s$ as function of $L$. It is obtained as the average value of $\theta_i$ in the middle half of the sites.
	}
\label{fig:figAppen2} %% label for entire figure
\end{figure}

In the following, we will focus on the spiral state, for which we have studied in some detail the finite-size effects on the spin configuration. In  Fig.~\ref{fig:figAppen2}(a) we show the same $q_s \simeq \pi/4$ state of Fig.~\ref{fig:figAppen1}(a), computed with a larger system size $L=200$. Besides boundary effects around $i=0,L$, there is a persisting modulation of the relative angle $\theta_i$, with amplitude $\delta\theta_i$. Further extending the system size (see inset), shows that the beating pattern does not shrink to zero in the thermodynamic limit, thus represents a genuine bulk feature of the system. Numerically, we find that the wavevector characterizing the beating pattern corresponds accurately to twice the Fermi wavevector of the chiral states, see Fig.~\ref{fig:fig1}(c). This clearly indicates the origin of this weak instability of the spiral state.

These long-wavelength modulation of $\theta_i$ can complicate the scaling analysis of the system. In Fig.~\ref{fig:figAppen2}(b) we plot the number of beating periods $N_B$ as function of system size $L$. For example, $N_B=5$ in panel~(a). As expected, $N_B$ grows almost linearly with system size. However, the growth of $N_B$ is due to regular jumps (with $\Delta N_B =2$), occurring for a change of length $\Delta L \simeq 60$. Between two jumps, the value of $N_B$ does not change. This behavior is easily understood noting that the length of the beating period is approximately $l_B \simeq 30$, see Fig.~\ref{fig:figAppen2}(a). Therefore, the spin configuration changes abruptly when two additional periods can be accommodated. In general, these jumps are generally associated with abrupt changes of other physical properties. For example, we see in Fig.~\ref{fig:figAppen2}(b) that the length $l_B$ increases gradually between two jumps, and (as expected) decreases abruptly when a larger system size can fit two additional beating periods. 
% Similar jumps are observed in the dependence of the numerically extracted spiral wavevector $q_s$ (see inset). At large $L$, the size of the jumps decreases, so $l_B$ and $q_s$ tend to well-defined limits.  

Finite-size effects are especially pronounced at smaller system sizes, and we find it useful to select values of $L$ which are approximately at the center of each $\Delta L \simeq 60$ interval. For example, we select the values $L= 70, 130, 190, \ldots$, marked by red dots in Fig.~\ref{fig:figAppen2}(b). By doing so, we can extract a smoother dependence of $l_B$ and other physical properties.

% We now provide an interpretation of the appearance of the beating pattern....

%\section{Existence of stationary and periodic solutions}

\section{Discrimination of dynamical phases}

\begin{figure}%[!htbp]
			\centering
			\includegraphics[width=0.47\textwidth]{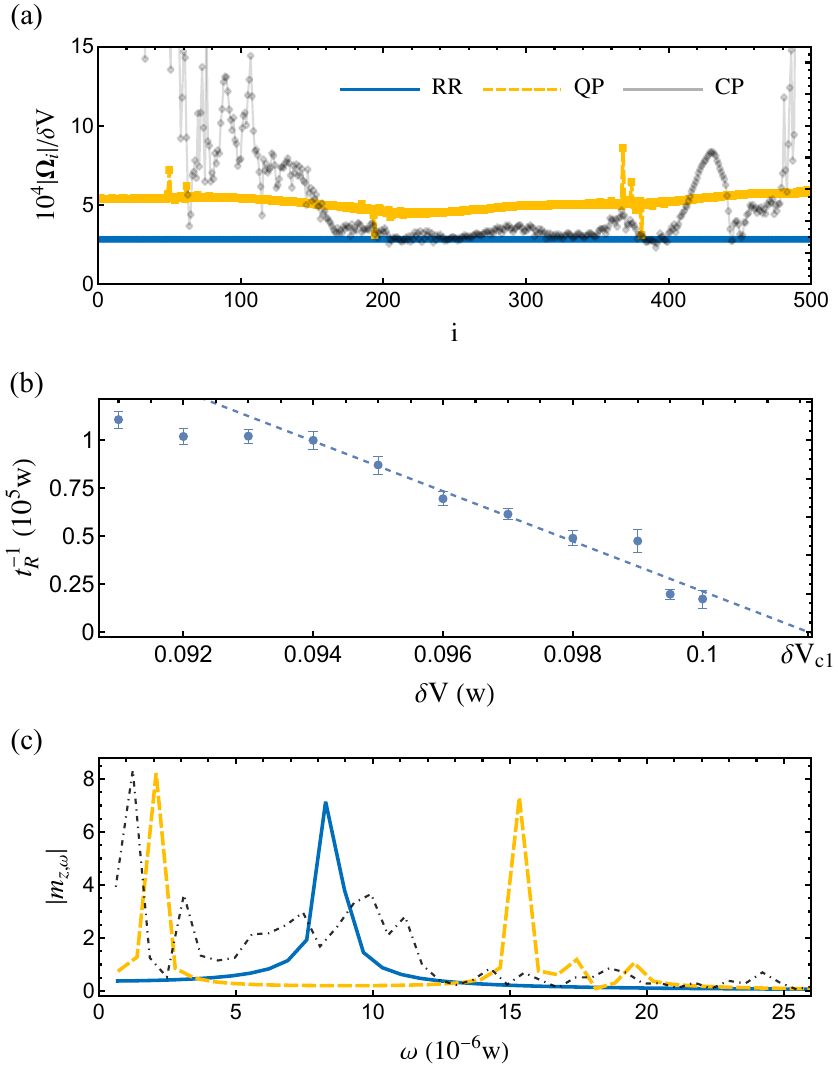}
	 \caption{(a) Values of $|\bm \Omega_i|$ for the RR (blue points), QP (orange points) and CP (gray points) states of Fig.~\ref{fig:comparison}, after evolving the system to $t=10^7w^{-1}$. (b) Bias dependence of the typical timescale to reach a uniform $\bm \Omega_i$. Each data point represents the average value of $t_R^{-1}$, obtained from 30 simulations with random initial states. The precise definition of $t_R$ is given in the main text. (c) Discrete-time Fourier transform of the time traces in Fig.~\ref{fig:comparison}(a). Explicitly, we compute $\vec m_{\omega}=(1/\sqrt{n})\sum_{j=0}^{n-1}	\vec m(t_j) e^{i\omega j}$, taking $n=1000$ equally spaced times $t_j$ in the interval $[10^6,10^7]w^{-1}$. In this figure the parameters are as in Fig.~\ref{fig:comparison}, except $L=30$ in panel (b). 
	}
\label{fig:figOmegai} %% label for entire figure
\end{figure}
%For each curve, a fixed random configuration is chosen for the initial state, to obtain a smooth dependence on $\delta V$. As seen, the extrapolated value $\delta V_{c1}$ of the phase boundary is independent of the initial state.

Generally, when solving Eq.~(\ref{m_evolution}) with increasing bias, the three different phases RR, QP, and CP appear sequentially. 
In the low-bias RR phase, the relative orientations of the $\bm m_i$ are locked and the spin configuration rigidly rotates in time. To determine numerically if a RR state has been established, we assume that neighboring sites $i$ and $i+1$ are rotating along the same axis. If this is the case, the common rotating vector $\vec\Omega_i$ is given by:
\begin{equation}
\vec\Omega_i = |\vec W_i|\frac{(\vec T_i \times \vec m_i) \cdot (\vec W_i \times \vec m_i)}{|\vec W_i \times \vec m_i|^2},
\end{equation}
where $\vec T_i = J\left \langle   \vec{{s}}_i \right \rangle+\eta (\left \langle   \vec{{s}}_i \right \rangle\times \vec{m}_i)$ and 
$\vec W_i = (\vec T_i \times \vec m_i) \times (\vec T_{i+1} \times \vec m_{i+1})$.
Therefore, the condition
\begin{equation}
\label{CriOfRR}
\vec\Omega_1=\vec\Omega_2=... =\vec\Omega_{L-1}
\end{equation} 
implies that all spins share a unique rotating vector $\bm \Omega_i =\bm \Omega$. In our numerical calculation, we decide that the system has reached a RR state when Eq.~(\ref{CriOfRR}) can be approximately satisfied, at the level $ \left(|\bm \Omega_i|_{max}-|\bm \Omega_i|_{min} \right)/|\bm \Omega_i|_{mean}<0.01$, after evolving the system for a long time, of the order $\sim 10 \times 2\pi/|\bm \Omega_{i}|_{mean}$. An explicit example is shown in Fig.~\ref{fig:figOmegai}(a), where $|\bm \Omega_i|$ is constant for the RR state of Fig.~\ref{fig:comparison} (blue dots) but depends on the site index $i$ for the QP state (orange dots) and CP state (gray dots).

In the RR phase, we also studied the timescale $t_R$ at which the $\bm \Omega_i$ approach a constant, finding that it diverges at the boundary of the QP phase. In practice, we define $t_R$ as the time at which the above condition of uniformity is met, i.e. $ \left(|\bm \Omega_i|_{max}-|\bm \Omega_i|_{min} \right)/|\bm \Omega_i|_{mean}<0.01$. The dependence of $t_R^{-1}$ on bias is shown in Fig.~\ref{fig:figOmegai}(b). We can get an accurate estimate of the critical bias $\delta V_{c1}$ between RR and QP by extrapolating the $t_R^{-1}$ data (dashed line). 

To identify the other two phases, QP and CP, we rely on the charge current and the explicit dynamics of the classical spins. Since the spin configuration of the QP and CP states does not simply precess rigidly, the charge current is time-dependent. Then, as shown in the main text, see Fig.~\ref{fig:comparison}(b), we can check if the charge current is periodic (QP) or displays a chaotic evolution (CP). Alternatively, we can distinguish the three phases from the dynamics of a selected spin, e.g., in the middle of the chain. By computing the discrete-time Fourier transform, we find a single peak (RR), several discrete peaks (QP), or a continuous frequency distribution (CP). In  Fig.~\ref{fig:figOmegai}(c) we show the  Fourier spectrum obtained from Fig.~\ref{fig:comparison}(a). 

\begin{figure}%[!htbp]
			\centering
			\includegraphics[clip, width=0.47\textwidth]{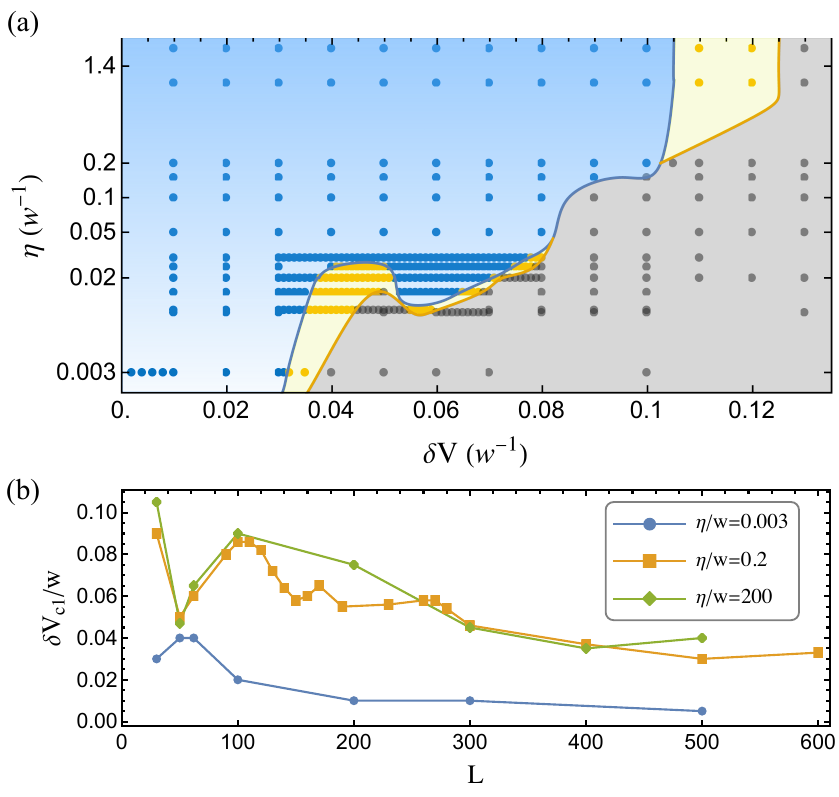}
	\caption{
	(a) Numerical phase diagram for $L=30$.
	(b) Finite size scaling of $\delta V_{c1}$ (the boundary between RR and QP) for $\eta/w=0.003,~0.2,200$.
	}
\label{fig:figAppen3} %% label for entire figure
\end{figure}

Using these methods, we have determined the phase diagram at a relatively large system size $L=500$, shown in Fig.~\ref{fig:phase_diagram} of the main text. We note, however, that the various phase boundaries depend on $L$.  An example of phase diagram at smaller system size ($L=30$) is shown in Fig.~\ref{fig:figAppen3}(a). Comparing it to Fig.~\ref{fig:phase_diagram}, we see that the details of the phase diagram depend on $L$ (e.g., the RR region is larger at smaller $L$), but the overall structure remains unchanged. With $L=30$, we have also checked that the system remains in the CP for values of $\delta V$ much larger than the range considered in Fig.~\ref{fig:figAppen3}(a). We further performed a finite-size scaling analysis of the RR-QP boundary at selected values of $\eta$, shown in Fig.~\ref{fig:figAppen3}(b), finding that the phase boundary has approximately converged when $L=500$. Therefore, for an infinite system the phase diagram would remain similar to Fig.~\ref{fig:phase_diagram}.

\begin{figure}%[!htbp]
			\centering
			\includegraphics[width=0.47\textwidth]{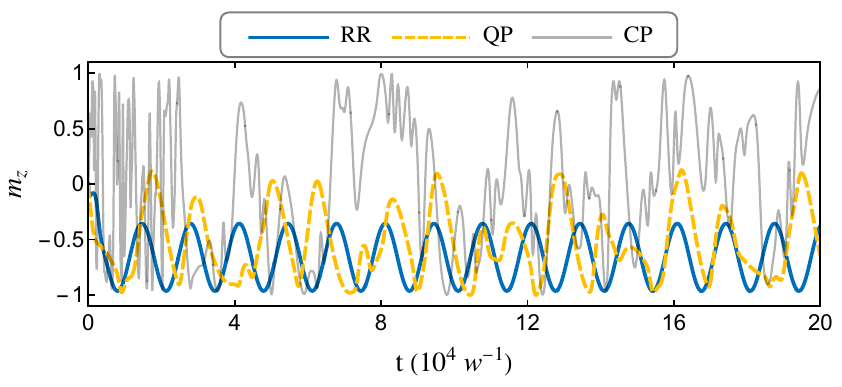}
	\caption{Evolution of $m_z$ for one of the two middle sites of a $L=30$ chain, with $J/w=0.1$, $\mu/w=-1.0$, and $ \eta/w =0.2 $. The bias is $ \delta V/w=0.2 $ (RR), $ 0.23$ (QP), and $0.51 $ (CP). 
		}
\label{fig:AnotherCase} %% label for entire figure
\end{figure}

Finally we note that our findings about the existence of three phases, RR, QP, and CP, as well as the general structure of the phase diagram, should remain qualitatively valid for a generic spiral state. In the main text we have selected the specific parameters $J/w = 1.4$ and $\mu/w = 1.5$. However, we find a similar behavior for other choices of $J,~\mu$. As an example, we consider in Fig.~\ref{fig:AnotherCase} the evolution of a small-$J$ spiral state ($J/w=0.1$, $\mu/w=-1.0$), under increasingly larger values of the bias $\delta V$. As seen, the dynamics is completely analogous to Fig.~\ref{fig:comparison}(a) of the main text.

\section{Spin current}
%\red{Charge current and spin current which are easily measured experimentally are calculated under adiabatic approximation. Charge current drained from left or right reservoirs to the chain is defined as $ \hat{J}_{cl}=-\partial _t \hat{N}_l $($l=L,R$) with $ \hat{N}_L=\sum_\sigma\sum_{i=-\infty }^{0}  c^\dagger_{i\sigma} c_{i\sigma} $ for $l=L$ and similarly for $\hat{N}_R$. Meir-Wingreen formula~\cite{PhysRevLett.68.2512} gives the relation between charge current and Green's function. Charge current can also be calculated inside the magnetic chain by $ J_{ci}=2w\sum_{\sigma} \rm{Im}(\left \langle c_{i\sigma}c^\dagger_{i+1\sigma} \right \rangle )$. $ J_{c}=J_{cL}=J_{ci}=J_{cR} $ guarantees charge current conservation of steady state. In the rigid-rotating phase, Fig.\ref{fig:comparison} shows that charge current is a constant value after some transient period as expected, i.e., $[J_c,H]=0$. }

\begin{figure}%[!htbp]
			\centering
			\includegraphics[width=0.47\textwidth]{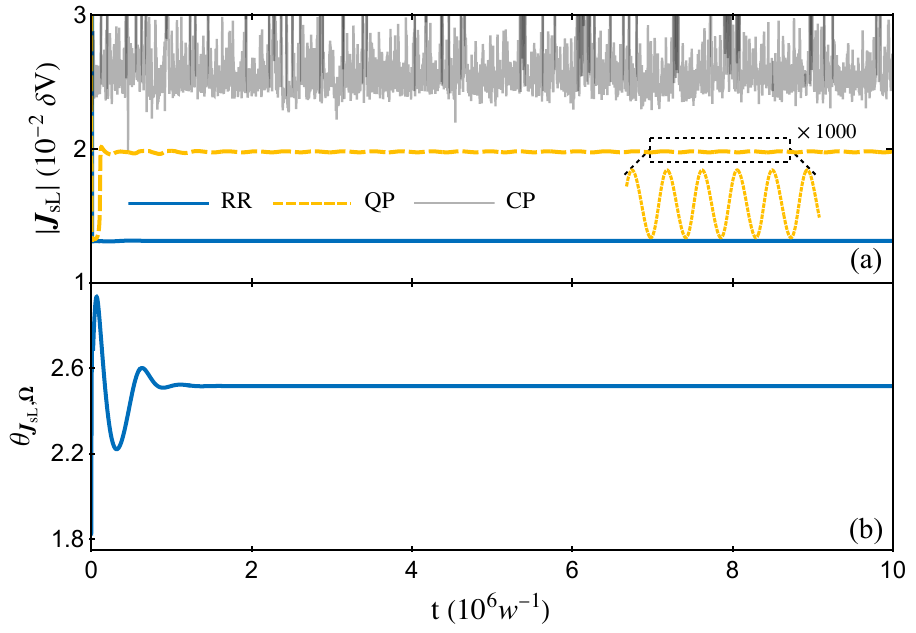}
	\caption{(a) Magnitude of the Spin current $|\vec J_{s\text{L}}|$. (b) Angle $\theta_{\vec J_{s\text{L}},\vec \Omega}$ between $\vec J_{s\text{L}}$ and $\vec \Omega$, as a function of time. We use the same parameters of Fig.~\ref{fig:comparison}.
	}
\label{fig:spinCurrent} %% label for entire figure
\end{figure}

While the charge current is constant for a RR state, detecting the spin current reveals its periodic time evolution. This has been shown explicitly in Fig.~\ref{fig:comparison}(c) of the main text, where the $z$ component is plotted. The oscillation of  $\vec J^z_{s \text{L}}$ is simply due to the fact that, in the RR phase, the spin current vector is rigidly rotating, with the same precession vector $\vec\Omega$. In Fig.~\ref{fig:spinCurrent} we show that $|\vec J_{s \text{L}}|$ and the angle between $\vec J_{s \text{L}}$ and $\vec\Omega$ become constant at sufficiently large times, when the RR has been established.

The spin current of Fig.~\ref{fig:comparison}(c) is defined as the rate of spin polarization drained from the left reservoir:
\begin{equation}\label{Js}
\vec J_{s \text{L}}= - \sum_{i=-\infty}^{0}\sum_{\sigma\sigma'}\partial_t\left \langle c^\dagger_{i\sigma}\vec{\sigma}_{\sigma\sigma'} c_{i\sigma'} ,\right \rangle.
\end{equation}
To compute $\vec J_{s \text{L}}$ we rely on the non-equilibrium Green's function formalism and express Eq.~(\ref{Js}) as a modified Meir-Wingreen formula~\cite{PhysRevLett.68.2512}: 
\begin{equation}
\vec J^\alpha_{s\text{L}}=\int_{\mu_{\rm R}}^{\mu_{\rm L}}\frac{d\omega}{2\pi} {\rm tr}\left[{\bm G}^R_c (\omega) {\bm \gamma}_{\rm R} {\bm G}^A_c (\omega){\bm \sigma}^\alpha{\bm \gamma}_{\rm L} \right],
\end{equation}
where ${\bm G}^R_c (\omega) = (\omega -\bm K)^{-1}$ is the retarded Green's function and ${\bm G}^A_c (\omega) = {\bm G}^R_c (\omega)^\dag$. The operators $\bm K$ and $\bm \gamma_{\rm L, R}$ are defined in the main text, shortly before Eq.~(\ref{chi_explicit}), and ${\bm \sigma}^\alpha$ are Pauli matrices. ${\bm \sigma}^\alpha$ only acts on the spin degrees of freedom, so its matrix elements are explicitly given as $[{\bm \sigma}^\alpha]_{i\sigma,i'\sigma'} = \delta_{i,i'}\sigma^\alpha_{\sigma,\sigma'}$. We can also define $\vec J_{s \text{R}}$ in an analogous manner, giving:
\begin{equation}
\vec J^\alpha_{s\text{R}}=\int_{\mu_{\rm R}}^{\mu_{\rm L}}\frac{d\omega}{2\pi} {\rm tr}\left[{\bm G}^R_c (\omega) {\bm \gamma}_{\rm L} {\bm G}^A_c (\omega){\bm \sigma}^\alpha{\bm \gamma}_{\rm R} \right].
\end{equation}
Since the spin current is not conserved, in general we have $\vec J_{s\text{L}} \neq \vec J_{s\text{R}}$. However, we find numerically that the magnitudes are nearly identical, i.e. $ |\vec J_{s\text{L}}|-|\vec J_{s\text{R}}| <10^{-6}$. 
%Furthermore, we find $|\vec J_{s\text{L,R}}| \simeq J_c$, although with a smaller degree of accuracy. 

Along the chain we can define the spin current as:
\begin{eqnarray}
\vec{J}^{i\to i+1}_s=2w\begin{pmatrix}
\rm{Im}(\chi_{i+1 \uparrow,i \downarrow} + \chi_{i+1 \downarrow,i \uparrow}) \\
\rm{Re}(\chi_{i+1 \uparrow,i \downarrow} +\chi_{i+1 \downarrow,i \uparrow}) \\
\rm{Im}(\chi_{i+1 \uparrow,i \uparrow}-\chi_{i+1 \downarrow,i \downarrow}) 
\end{pmatrix},
\end{eqnarray}
satisfying $\vec{J}^{i-1\to i}_s-\vec{J}^{i\to i+1}_s=J\left \langle \vec s_i \right \rangle \times \vec m_i$. We see that the contribution from classical spins' magnetic moments has to be taken into account to get a conservation formula for the spin current. At site $i$, the difference between the incoming and outgoing spin current is exactly the torque acting on the localized moment which, without damping, also equals $\partial_t\vec m_i$.

\section{Infinite system evolution}

To treat the infinite system, we apply the gauge transformation of Eq.~(\ref{GTransf}) and solve the equation of motion:
\begin{equation}\label{tilde_m_evolution}
\frac{\partial \bar{\bm m}_i }{\partial t} =J\left \langle   \vec{{s}}_i \right \rangle\times \bar{\bm m}_i+\eta (\left \langle   \vec{{s}}_i \right \rangle\times \bar{\bm m}_i)\times \bar{\bm m}_i,
\end{equation}
where $\bar{\bm m}_i$ are the local magnetic moments in the transformed frame. The electron spin polarization $ \left \langle \vec{{s}}_i \right \rangle $ should now be computed based on $\bar{H}_c$, see Eq.~(\ref{tildeHc}), instead of $H_c$. The gauge transformation is useful as it brings the conical states (\ref{tilted_spiral}) to uniform configurations of the localized moments, i.e., $\bar{\bm m}_i$ does not depend on the site index. As a consequence, the local spin polarization $ \left \langle \vec{{s}}_i \right \rangle $ is also independent of $i$ and the time evolution determined by Eq.~(\ref{tilde_m_evolution}) is identical for all the  $\bar{\bm m}_i$.

Since the $\bar{\bm m}_i$ remain uniform, not only Eq.~(\ref{tilde_m_evolution}) can be solved for a single site, but also we can easily obtain the instantaneous electronic bands $E_\pm (k)$ and the corresponding eigenstates.  The spin polarization $ \left \langle \vec{{s}}_i \right \rangle $ is simply determined by the occupation numbers $n_\pm (k)$. In our case, the $-$ band is always fully occupied while $n_+ (k)$ has the general double-step structure~\cite{PhysRevB.96.054302}:
\begin{equation}
n_{+}(k)=\left\{
\begin{array}{lc}
 0 ~~~&k \leq k_{\text{L}-}, \\
 b_- &k_{\text{L}-}<k \leq k_{\text{R}-}, \\
 1 &k_{\text{R}-}<k \leq k_{\text{R}+}, \\
  b_+&k_{\text{R}+}<k \leq k_{\text{L}+}, \\
  0&k>k_{\text{L}+},
\end{array}\right.
\end{equation}
where $k_{\alpha\pm}$ are the solutions of $E_+(k)=\mu_{\alpha}$ ($\alpha = {\rm L,R}$ and $k_{\alpha-}< 0 < k_{\alpha+}$). A main difficulty remains, as the values of $b_\pm$ are determined by scattering at the contacts, thus cannot be computed based on the infinite model alone. 

\begin{figure}%[!htbp]
			\centering
			\includegraphics[width=0.47\textwidth]{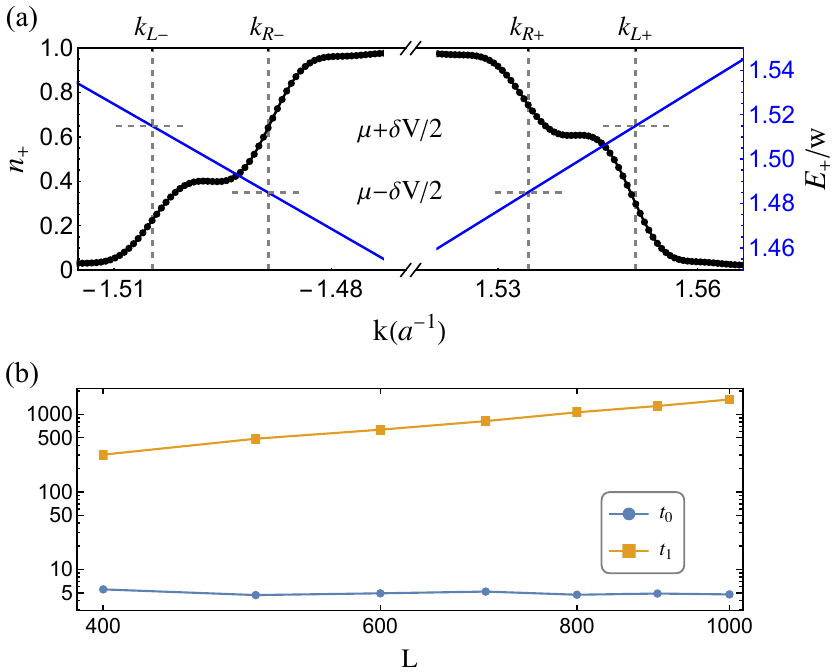}
	\caption{(a) Distribution function $n_{+}(k)$ (black dots), extracted at $t=500w^{-1}$ for a system size $L=1000$, and used to simulate dynamics of an infinite system in Fig.~\ref{fig:InfiniteSystem}. We also show the dispersion relation $E_+(k)$ (solid blue lines). (b) Dependence of $t_0$ [see Eq.~(\ref{t0_def})] and $t_1$ [see Eq.~(\ref{t1_def})]  on the system size $L$. We have used the parameters of Fig.~\ref{fig:InfiniteSystem}.
	}
\label{fig:t1t0} %% label for entire figure
\end{figure}

To extract the occupation function we rely on:
\begin{eqnarray}
	n_{\alpha}(k)&=&\left \langle d_{k\alpha}^\dagger d_{k\alpha} \right \rangle \nonumber\\
		&=&\frac{1}{L}\sum_{j,j'}\sum_{\sigma\sigma'}S_{\alpha\sigma}^*(k) S_{\alpha\sigma'}(k)  e^{ik(j-j')}\nonumber\\
		&&(\delta_{j,j'}\delta_{\sigma,\sigma'}-\chi_{j'\sigma',j\sigma}),
\end{eqnarray}
which can be easily obtained from the expression of the eigenmodes:
\begin{equation}
d_{k\alpha} =\frac{1}{\sqrt{L}} \sum_{j,\sigma} S_{\alpha \sigma}(k) e^{-i k j} c_{j\sigma}.
\end{equation}
The $2\times 2$ unitary matrix $\bm{S}(k)$ can be obtained from Eq.~(\ref{H0_matrix}) and we perform finite-size simulations (with large $L$), to compute $\bm{\chi}$ using Eq.~(\ref{chi_explicit}). We show in Fig.~\ref{fig:t1t0} the occupation function obtained in this manner, corresponding to the $L=1000$ simulation of Fig.~\ref{fig:InfiniteSystem}. We observe the expected double-step structure, where the discontinuities are broadened due to the finite system size. This allows us to obtain accurate estimates of $b_\pm$ and compute the infinite-system dynamics shown in Fig.~\ref{fig:InfiniteSystem} (red dashed curve).

As seen in Fig.~\ref{fig:InfiniteSystem}, this procedure can reproduce well the finite-size simulations. Some slight deviations appear at short times, which are probably due to a small change in the distribution function. In principle, since the spin texture changes in time, the values $b_\pm$ should also be taken as time-dependent. Instead, we have extracted $n_{+}$ at a fixed time, $t= 500 w^{-1}$, corresponding to the metastable tilted spiral state. In this regime when the $m_{i,z}$ have saturated to a constant value, we indeed find excellent agreement between the red dashed curve and the finite-size simulation. Notice, however, that we always have $m_{j,z} \ll 1 $ in Fig.~\ref{fig:InfiniteSystem}, indicating that the spin texture is only weakly deformed from the initial planar spiral. Thus, we expect that also the $b_\pm$ are almost constant in time, which is confirmed by the generally good agreement found.

Finally, we analyze two timescales characterizing the dynamics, $t_0$ and $t_1$, which in Fig.~\ref{fig:t1t0}(b) are computed for the system of Fig.~\ref{fig:InfiniteSystem}. The value of $t_0$ is given by:
\begin{equation}\label{t0_def}
\overline{m_{i,z}(t_0)} =  f \times \overline{ m_{i,z}(t_{ref})},
\end{equation}
while $t_1$ is obtained as follows:
\begin{equation}\label{t1_def}
{\rm max}_i \left|m_{i,z}(t_1) - \overline{ m_{i,z}(t_{ref})}\right| = \epsilon,
\end{equation}
where the index $i$ is restricted to the bulk of the chain, i.e., labels the $40\%$ middle spins considered in Fig.~\ref{fig:InfiniteSystem}, and the overline indicates the average over them. The reference time $t_{ref}$ is in the metastable regime and, specifically,  we have used $t_{ref}=100 w^{-1}$ in Fig.~\ref{fig:t1t0}(b). $f$ is a fraction close to 1 (we take $f=90\%$) and $\epsilon$ is a small error (we take $\epsilon=0.01$). While $t_0$ characterizes the transient of the time evolution, $t_1$ is the time at which the infinite-system approximation breaks down. In the regime $t_0 < t < t_1$ the system is in a metastable tilted spiral, thus $m_{z,i}$ is approximately independent of $i$ and constant in time.  

In Fig.~\ref{fig:t1t0}(b) we show the dependence of $t_0,t_1$  on system size at relatively large values of $L$. We see that, as expected, $t_0$ is approximately constant, reflecting the convergence of the transient dynamics with $L$. Instead, the value of $t_1$ grows with $L$, implying that the metastable conical state has an infinite lifetime in the thermodynamic limit. As we have discussed in the main text, at any finite value of $L$ the system will eventually display a nontrivial dynamics when $t\to \infty$. In the case of Fig.~\ref{fig:InfiniteSystem}, the model will settle into a RR state, with a period growing as $L^3$. Therefore, we can interpret $t_1$ as the timescale at which the RR state starts to take form. Obviously, however, the periodic motion has not yet been fully established (since at $t=t_1$ the $\bm{m}_i$ are still relatively close to the metastable configuration). From Fig.~\ref{fig:t1t0}(b) we find $t_1 \propto L^{\alpha}$, with $\alpha \simeq 1.75$, which is consistent with this interpretation. The precise value of $\alpha$ might be affected by details in the definition of $t_1$, e.g., the choice of $\epsilon$ and the set of bulk spins, but the exponent must be compatible with the dependence $|\bm{\Omega}| \propto L^{-3}$. Indeed, we find $\alpha < 3$.

%\bibliography{nonequ.bib}
\end{document}